\documentstyle[12pt]{article}
\catcode`\@=11

\@addtoreset{equation}{section}
\def\theequation{\thesection\arabic{equation}}

\def\@normalsize{\@setsize\normalsize{15pt}\xiipt\@xiipt
\abovedisplayskip 14pt plus3pt minus3pt%
\belowdisplayskip \abovedisplayskip
\abovedisplayshortskip  \z@ plus3pt%
\belowdisplayshortskip  7pt plus3.5pt minus0pt}
\def\small{\@setsize\small{13.6pt}\xipt\@xipt
\abovedisplayskip 13pt plus3pt minus3pt%
\belowdisplayskip \abovedisplayskip
\abovedisplayshortskip  \z@ plus3pt%
\belowdisplayshortskip  7pt plus3.5pt minus0pt
\def\@listi{\parsep 4.5pt plus 2pt minus 1pt
            \itemsep \parsep
            \topsep 9pt plus 3pt minus 3pt}}

\def\underline#1{\relax\ifmmode\@@underline#1\else
        $\@@underline{\hbox{#1}}$\relax\fi}
\@twosidetrue
\relax

\catcode`@=12

\catcode`\@=11

\def\section{\@startsection{section}{1}{\z@}{3.5ex plus 1ex minus
   .2ex}{2.3ex plus .2ex}{\large\bf}}
\def\thesection{\arabic{section}.}

\def\ps@headings{\def\@oddfoot{}\def\@evenfoot{}
\def\@oddhead{\hbox{}\hfill
        \makebox[.5\textwidth]{\raggedright\ignorespaces --\thepage{}--
        \hfill }}
\def\@evenhead{\@oddhead}
\def\subsectionmark##1{\markboth{##1}{}}
}

\ps@headings

\catcode`\@=12

\relax

\def\figcap{\section*{Figure Captions\markboth
        {FIGURECAPTIONS}{FIGURECAPTIONS}}\list
        {Fig. \arabic{enumi}:\hfill}{\settowidth\labelwidth{Fig. 999:}
        \leftmargin\labelwidth
        \advance\leftmargin\labelsep\usecounter{enumi}}}
 \relax
\def\tablecap{\section*{Table Captions\markboth
        {TABLECAPTIONS}{TABLECAPTIONS}}\list
        {Table \arabic{enumi}:\hfill}{\settowidth\labelwidth{Table 999:}
        \leftmargin\labelwidth
        \advance\leftmargin\labelsep\usecounter{enumi}}}
 \relax
\def\reflist{\section*{References\markboth
        {REFLIST}{REFLIST}}\list
        {[\arabic{enumi}]\hfill}{\settowidth\labelwidth{[999]}
        \leftmargin\labelwidth
        \advance\leftmargin\labelsep\usecounter{enumi}}}
 \relax

\catcode`\@=11

\def\marginnote#1{}
\newcount\hour
\newcount\minute
\newtoks\amorpm
\hour=\time\divide\hour by60
\minute=\time{\multiply\hour by60 \global\advance\minute by-
\hour}
\edef\standardtime{{\ifnum\hour<12 \global\amorpm={am}%
    \else\global\amorpm={pm}\advance\hour by-12 \fi
    \ifnum\hour=0 \hour=12 \fi
    \number\hour:\ifnum\minute<100\fi\number\minute\the\amorpm}}
\edef\militarytime{\number\hour:\ifnum\minute<100\fi\number\minute}
\def\draftlabel#1{{\@bsphack\if@filesw {\let\thepage\relax
  \xdef\@gtempa{\write\@auxout{\string
    \newlabel{#1}{{\@currentlabel}{\thepage}}}}}\@gtempa
    \if@nobreak \ifvmode\nobreak\fi\fi\fi\@esphack}
     \gdef\@eqnlabel{#1}}
\def\@eqnlabel{}
\def\@vacuum{}
\def\draftmarginnote#1{\marginpar{\raggedright\scriptsize\tt#1}}
\def\draft{\oddsidemargin -.5truein
        \def\@oddfoot{\sl preliminary draft \hfil
        \rm\thepage\hfil\sl\today\quad\militarytime}
        \let\@evenfoot\@oddfoot \overfullrule 3pt
        \let\label=\draftlabel
        \let\marginnote=\draftmarginnote
   
\def\@eqnnum{(\theequation)\rlap{\kern\marginparsep\tt\@eqnlabel}%
\global\let\@eqnlabel\@vacuum}  }
\def\preprint{\twocolumn\sloppy\flushbottom\parindent 1em
        \leftmargini 2em\leftmarginv .5em\leftmarginvi .5em
        \oddsidemargin -.5in    \evensidemargin -.5in
        \columnsep 15mm \footheight 0pt
        \textwidth 250mmin      \topmargin  -.4in
        \headheight 12pt \topskip .4in
        \textheight 175mm
        \footskip 0pt
        
\def\@oddhead{\thepage\hfil\addtocounter{page}{1}\thepage}
        \let\@evenhead\@oddhead \def\@oddfoot{} \def\@evenfoot{} 
}
\def\titlepage{\@restonecolfalse\if@twocolumn\@restonecoltrue\onecolumn
     \else \newpage \fi \thispagestyle{empty}\c@page\z@
        \def\thefootnote{\fnsymbol{footnote}} }
\def\endtitlepage{\if@restonecol\twocolumn \else  \fi
        \def\thefootnote{\arabic{footnote}}
        \setcounter{footnote}{0}}  
\catcode`@=12
\relax

\def\ps@headings{\def\@oddfoot{}\def\@evenfoot{}
\def\@oddhead{\hbox{}\hfill
        \makebox[.5\textwidth]{\raggedright\ignorespaces --\thepage{}--
        \hfill }}
\def\@evenhead{\@oddhead}
\def\subsectionmark##1{\markboth{##1}{}}
}

\ps@headings

\relax

\def\firstpage#1#2#3#4#5#6{
\begin{document}
\begin{titlepage}
\nopagebreak
\title{\begin{flushright}
        \vspace*{-1.8in}
        {\normalsize CERN-TH/97-249}\\[-9mm]
        {\normalsize hep-th/9710023}\\[4mm]
\end{flushright}
\vspace{3cm}
{#3}}
\author{\large #4 \\[0.0cm] #5}
\maketitle
\vskip 5mm
\nopagebreak 
\begin{abstract}
{\noindent #6}
\end{abstract}
\vfill
\begin{flushleft}
\rule{16.1cm}{0.2mm}\\[-3mm]
$^{\star}${\small Research supported in part by\vspace{0mm}
 the EEC under TMR contract ERBFMRX-CT96-0090.
\\[-3mm]
$^\dag$e-mail: Alexandros.Kehagias@cern.ch \\[-3mm]
\hspace{1.45cm}Herve.Partouche@cern.ch } \\[0mm]
CERN-TH-97-249\\
October 1997
\end{flushleft}
\thispagestyle{empty}
\end{titlepage}}

\def\simlt{\stackrel{<}{{}_\sim}}
\def\simgt{\stackrel{>}{{}_\sim}}
\newcommand{\dal}{\raisebox{0.085cm}
{\fbox{\rule{0cm}{0.07cm}\,}}}

\newcommand{\be}{\begin{equation}}
\newcommand{\ee}{\end{equation}}
\newcommand{\btau}{\bar{\tau}}
\newcommand{\p}{\partial}
\newcommand{\bp}{\bar{\partial}}

\newcommand{\gsi}{\,\raisebox{-0.13cm}{$\stackrel{\textstyle
>}{\textstyle\sim}$}\,}
\newcommand{\lsi}{\,\raisebox{-0.13cm}{$\stackrel{\textstyle
<}{\textstyle\sim}$}\,}
\date{}
\firstpage{3118}{IC/95/34}
{\large {\Large O}N THE 
{\Large E}XACT {\Large Q}UARTIC {\Large E}FFECTIVE {\Large A}CTION 
\\
FOR THE {\Large T}YPE {\Large IIB} {\Large S}UPERSTRING$^\star$ \\
\phantom{X}}
{A. Kehagias and H. Partouche$^\dag$} 
{
\normalsize\sl Theory Division, CERN, 1211 Geneva 23, Switzerland
}
{ We propose a four-point effective action  
for the graviton, antisymmetric two-forms, dilaton and  axion of type IIB 
superstring in ten dimensions. It is explicitly 
$SL(2,\bf{Z})$-invariant and 
reproduces the  known tree-level results.  Perturbatively, it 
has  only one-loop corrections for the NS-NS sector, generalizing 
the non-renormalization theorem of the $R^4$ 
term.  Finally, the non-perturbative 
corrections are of the expected form, namely,  they    
can be interpreted as arising from single D-instantons 
of multiple charge.}
\section{Introduction}

There is a lot of activity nowadays towards understanding the 
non-perturbative  structure in string theory. 
In type IIB theory, in particular,  the non-perturbative physics is 
intimately  related with the  
existence of the $SL(2,\bf{Z})$ symmetry \cite{S,HT}.  
The spectrum of the type IIB 
theory contains  an 
$SL(2,\bf{Z})$ multiplet of strings  and five-branes, the 
self-dual three-brane, the seven-brane, as well as D-instanton solutions.
The latter are the only ones that give non-perturbative corrections
in ten dimensions. 
This can be seen by compactifying the theory. 
In this case, the various  Euclidean $(p+1)$-word-volumes of p-branes have an 
infinite action in the decompactification limit except when $p=-1$, which is 
just the type IIB D-instanton. 

In $\sigma$-model perturbation theory, there exists a four-loop divergence 
that contributes to the $\beta$-functions \cite{G} and gives 
 $\alpha'^3$ corrections to  the effective action. 
This can also be confirmed by string four-point amplitude  calculations 
\cite{S2,GS}. 
For four gravitons, in particular, there exists also a one-loop result
\cite{ST}
for  the $R^4$ corrections and non-renormalization theorems have been 
conjectured for their structure \cite{T,GG}. One expects that
 all  contributions higher than one loop to vanish, since for higher 
genus surfaces there are more than eight fermionic zero modes; this is 
 exactly the number needed to saturate the 
external particles in a four-point amplitude \cite{GG}. This heuristic 
argument has been proved by using superspace techniques \cite{B}.

Besides  the perturbative corrections to the $R^4$ term, there also exist
non-perturbative ones. Their form has recently been conjectured by Green and 
Gutperle on the basis of $SL(2,\bf{Z})$ invariance \cite{GG}. In particular, 
the modular invariance of the effective action is achieved by employing 
a certain non-analytic modular form. The structure of the latter is such 
that it gives only tree- and one-loop corrections to the $R^4$ term besides 
the instanton ones. An ansatz for the form of the corresponding 
four-graviton amplitude 
has been given in \cite{R}. Moreover, 
the $R^4$ term gives rise to a similar term 
in M-theory \cite{GGV,GV,RT} and the
compactification of the latter gives results consistent with string
theory expectations \cite{Str,ANT}. 
  
One may now proceed further by including the other massless modes of 
the type IIB theory. In this case, the tree-level result for  the  
four-point amplitudes of the dilaton and the antisymmetric tensor 
has been given in \cite{GS},
while  a one-loop calculation  is lacking. Now, arguments similar to those 
above  
seem to suggest that the non-renormalization theorem for the $R^4$ term 
may also be extended to the full effective theory when all modes are included.
Namely,  the perturbative expansion for the NS-NS sector 
stops at one loop and all 
other corrections are non-perturbative. 
This can also be justified by consistency conditions related to M-theory 
\cite{KP1}.
However, the inclusion of the other modes  
at the tree level has a serious drawback. It breaks the 
manifest $SL(2,\bf{Z})$ invariance of the theory. Here, we propose 
an effective action for all  bosonic massless modes of type IIB, except 
for the  self-dual four-form. We do not consider the latter
 because of the lack of any perturbative information 
at the eight-derivative level.
The action  we propose   respects the $SL(2,\bf{Z})$ symmetry and 
reproduces the effective action of \cite{GS} when all R-R fields are
switched off. In particular, the NS-NS sector 
has only tree- and one-loop corrections 
besides the non-perturbative ones. 

In the following section,  we recall perturbative  results in 
the type IIB effective theory. In section 3, we summarize the analysis 
of \cite{GG} concerning the non-perturbative corrections to the $R^4$ term. 
In section 4, we propose an 
$SL(2,\bf{Z})$-invariant effective action  and discuss its 
compatibility with a recent calculation \cite{APT} of the $R^2(\p\p\phi)^2$ 
term in type IIB on $K3$.

\section{Perturbative Effective Type IIB Theory}

The massless  bosonic spectrum of  type IIB superstring theory 
consists in  the graviton $g_{MN}$, 
the dilaton $\phi$ and the antisymmetric tensor $B^1_{MN}$ 
in the NS-NS sector and the axion  $\chi$, the two-form $B^2_{MN}$ and the 
self-dual four-form field $A_{MNPQ}$ in the R-R sector. The two scalars of 
the theory can be combined into a complex 
one, $\tau=\tau_1+i\tau_2$, defined by
\be
\tau=\chi+ie^{-\phi}\, . \label{tau}
\ee
The theory has two 
supersymmetries generated by two supercharges of the same chirality.   
It has in addition a conserved $U(1)$ 
charge which generates  rotations of the two 
supersymmetries and under  which some of the  fields  are charged 
\cite{S}. 
The graviton and the four-form field are neutral,  the antisymmetric  
tensors have charge  $q=1$, whereas the complex scalar $\tau$ has $q=2$.
The fermionic superpartners of the above fields are a complex Weyl gravitino 
and a  complex Weyl dilatino.  

The bosonic effective Lagrangian of the theory in lowest order in 
$\alpha'$ takes the form\footnote{We  set $\alpha'=1$ from now on.}
\be
{\cal{L}}_{0}= 
R-\frac{1}{2\tau_2^2}\partial_M\tau\partial^M\btau
-\frac{1}
{12\tau_2}(\tau H^1+H^2)_{KMN}(\btau H^1+H^2)^{KMN}\, , 
\label{IIB}
\ee
where $H^\alpha_{KMN}=\p_KB^\alpha_{MN}+\mbox{cyclic}$ for $\alpha=1,2$ 
and we have set the four-form to zero. The theory has an 
$SL(2,\bf{R})$ symmetry that acts as 
\be
\tau\rightarrow \frac{a\tau+b}{c\tau+d}\, , ~~~~
B^\alpha_{MN}\rightarrow {(\Lambda^T)^{-1}}^\alpha_{\mbox{\phantom{a}}\beta} 
B^\beta_{MN} 
\, , ~~~ \Lambda= \left(\matrix{a & b 
\cr c& d}\right) \in SL(2,\bf{R})\, ,
\ee
and leaves the  Lagrangian (\ref{IIB}) invariant.
The complex scalar $\tau$ parametrizes an $SL(2,{\bf R})/U(1)$ coset space. 
In general, the group 
$SL(2,\bf{R})$ can be represented by a matrix $V^\alpha_{\pm}$ 
\cite{S,GG}
\be
 V=\left(\matrix{V^1_- & V^1_+\cr 
V^2_- & V^2_+ }\right)= \frac{1}{\sqrt{-2 i\tau_2}} 
\left(\matrix{\btau e^{-i\theta} & \tau e^{i\theta} \cr 
e^{-i\theta} & e^{i\theta} }\right)\, . 
\ee
 The local $U(1)$ is realized by the shift 
$\theta\rightarrow \theta+\Delta\theta$ and the global 
$SL(2,\bf{R})$ acts from the left. One may define the quantities 
\be
P_M=-\epsilon_{\alpha\beta}V^\alpha_+\p_MV^\beta_+=ie^{2i\theta}
\frac{\p_M\tau}{2\tau_2}\, , ~~~ Q_M=-i\epsilon_{\alpha\beta}V^\alpha_+
\p_MV^\beta_-=\p_M\theta -
\frac{\p_M\tau_1}{2\tau_2}\, ,   \label{PQ} 
\ee
where $Q_M$ is a composite $U(1)$ gauge connection and $P_M$ has charge 
$q=2$. We also define the complex three-form
\be
G_{KMN}=-\sqrt{2i}\delta_{\alpha\beta}
V^\alpha_+H^\beta_{KMN}=-i\frac{e^{i\theta}}{\sqrt{\tau_2}}
(\tau H^1_{KMN}+H^2_{KMN})\, , \label{G}
\ee
with charge $q=1$. We  fix the gauge by choosing  
$\theta\equiv 0$ from now on. 
In this case, the global $SL(2,\bf{R})$ transformation is 
non-linearly realized and the various quantities in eqs.(\ref{PQ}) and 
(\ref{G}) transform as
\be
P_M\rightarrow \frac{c\btau+d}{c\tau+d}P_M\, , ~~~ Q_M\rightarrow 
Q_M+\frac{1}{2i}\p_M\ln \left(\frac{c\btau+d}{c\tau+d}\right)\, ,~~~
G_{KMN}\rightarrow \left(\frac{c\btau+d}{c\tau+d}\right)^{1/2} G_{KMN}
\, . \label{tra}
\ee
We may also define  the covariant derivative $D_M=\nabla_M-iqQ_M$, 
which transforms under $SL(2,\bf{R})$ as  
\be
D_M\rightarrow \left(\frac{c\btau+d}{c\tau+d}\right)^{q/2}D_M \, . \label{D} 
\ee

There exists  ${\alpha'}^3$ corrections to
the  effective Lagrangian (\ref{IIB})
above, which have  been evaluated in \cite{GS} and are written as
\begin{eqnarray}
{\cal{L}}_{4pt}&=&\frac{\zeta(3)}{3\cdot 2^6}\tau_2^{3/2}
\Big{(}t_8^{ABCDEFGH}t_8^{MNPQRSTU}+
\frac{1}{8}\varepsilon_{10}^{ABCDEFGHIJ}{\varepsilon_{10}^{MNPQRSTU}}_{IJ}
\Big{)} \nonumber \\
&&
\times \hat{R}_{ABMN}\hat{R}_{CDPQ}\hat{R}_{EFRS}\hat{R}_{GHTU} \, , \label{R4}
\end{eqnarray}
where  
\be
{\hat{R}_{MN}}^{\mbox{\phantom{PQ}}PQ}={R_{MN}}^{PQ}+\frac{1}{2}
e^{-\phi/2}\nabla_{[M}{H^1_{N]}}^{PQ}-\frac{1}{4}
{g_{[M}}^{[P}\nabla_{N]}\nabla^{Q]}\phi ~~.  \label{hR}
\ee
The tensor $t_8$ is defined in \cite{S2}, $\varepsilon_{10}$ is the totally
antisymmetric symbol in ten dimensions and
the square brackets are defined without the  combinatorial factor 
$1/2$ in front.
The Lagrangian (\ref{R4}) reproduces the four-point amplitude  
calculated in string theory and it is in agreement  with  
$\sigma$-model perturbative calculations. 
In addition, at the same order in $\alpha'$ one expects contributions 
coming from five- to eight-point amplitudes. These contributions can be 
implemented by adding the terms 
\begin{eqnarray}
&&-\frac{1}{4}e^{-\phi}
{H^1_{[M}}^{C[P}{H^1_{N]C}}^{Q]}+\frac{1}{16}{g_{[M}}^{[P}\p_{N]}\phi
\p^{Q]}\phi-\frac{1}{8}{g_{[M}}^{[P}{g_{N]}}^{Q]}\p_K\phi\p^K\phi
 \label{hR1}\\
&&-\frac{1}{4}e^{-\phi/2}\p^{[P}\phi{{H^1}^{Q]}}_{MN}-{1\over 4}e^{-\phi/2}
\partial_{[M}\phi {H^1_{N]}}^{PQ}-{1\over 8}e^{-\phi/2}{g_{[M}}^{[P}
{H^1_{N]}}^{Q]C}\partial_C\phi  
   \nonumber
\end{eqnarray}
to the r.h.s. of  eq.(\ref{hR})  and then $\hat{R}_{PQMN}$ turns out to be 
the  Riemann tensor 
associated to the  generalized connection, which includes the torsion and the
Weyl connection \cite{GS}
\be
T_{MNP}=\frac{1}{2}e^{-\phi/2}H^1_{MNP}\, , ~~~
\tilde{\omega}_{MNP}=\frac{1}{4}g_{P[N}\p_{M]}\phi\, . 
\ee
Expanding the $\hat{R}$ terms in 
eq.(\ref{R4}) we find  
\begin{eqnarray}
{\cal{L}}_{4pt}&=&\frac{\zeta(3)}{3\cdot 2^6}\tau_2^{3/2}
\left(t_8t_8+\frac{1}{8}\varepsilon_{10}\varepsilon_{10}\right)
\left(R^4+e^{-2\phi}(\nabla H^1)^4+(\p\p\phi)^4
\right. \nonumber \\
&&-
12e^{-\phi}R(\nabla H^1)^2\p\p\phi 
-4R^3\p\p\phi+6e^{-\phi}R^2(\nabla H^1)^2 \label{LL} \\ &&\left.
+6e^{-\phi}(\nabla H^1)^2(\p\p\phi)^2
+6R^2(\p\p\phi)^2-4R(\p\p\phi)^3\right)\, , \nonumber 
\end{eqnarray}
where $\p\p\phi$ stands for $(\p\p\phi)_{MNPQ}\equiv g_{MP}\nabla_N\p_Q\phi$ 
and $\nabla H^1$ for $(\nabla H^1)_{MNPQ}\equiv \nabla_M H^1_{NPQ}$. 
It should be noted 
that there are no terms with odd powers of $\nabla H^1$, because such  
terms vanish, owing to the Bianchi identity. One 
can check for example that the amplitude for three gravitons and 
one antisymmetric field
\be
t_8^{ABCDEFGH}t_8^{MNPQRSTU}
R_{ABMN}R_{CDPQ}R_{EFRS}\nabla_{[G}H^1_{H]TU}  
\ee
is zero. This can also be argued on the basis of the invariance of the type 
IIB superstring under the world-sheet parity that acts on $B^1$ as $B^1\to 
-B^1$.  

At the perturbative level, there exist string one-loop  corrections 
to the four-point functions. 
For four gravitons these corrections have been calculated 
\cite{ST} and  amount to the exchange 
\be
\zeta(3)\rightarrow \zeta(3)+\frac{\pi^2}{3}\tau_2^{-2}\, , 
\ee
in eq.(\ref{R4}).   

\newpage 

\section{Non-perturbative $R^4$ couplings}

In the type IIB theory, there exist D-instantons that contribute to the 
four-point amplitudes. Instantons are solutions of the tree-level Lagrangian 
(\ref{IIB}) in Euclidean space with vanishing antisymmetric fields and 
non-trivial profile for the complex scalar $\tau$ \cite{GGP}. 
In general, they break half of the supersymmetries and the broken 
ones generate fermionic zero modes. Since the supersymmetry in type IIB 
theory  is generated by a complex Weyl spinor with 16 components, we expect
eight fermionic zero modes, which  can  give a non-zero 
contribution to  four-point amplitudes. In such a background, the action 
is finite and its value is $S^{(Q)}=-2\pi|Q| i\tau_0$, where $Q$ is the 
instanton charge and $\tau_0$ is the value of $\tau$ at infinity. 
Thus, we expect the  contribution  from 
a single instanton of charge 1 to be proportional 
to $e^{2\pi i \tau_0}$.  The multi-instanton contributions 
may be determined by T-duality arguments as follows \cite{GG}.  
Under compactification on $S^1$, the type IIB D-instanton 
is mapped to the type IIA D-particle. There are arguments to support the 
fact that $n$ such single charged D-particles combine to a 
single bound state  of charge $n$ \cite{W}. 
The world line of this bound state can wrap
$m$ times around the compact $S^1$  so that its topological charge is $mn$.
Then, its T-dual counterpart in type IIB should be a D-instanton
of  charge $Q=mn$ whose  
contribution is proportional to $e^{2i\pi |mn|\tau_0}$.
Notice that  separated instantons are 
accompanied by additional  fermionic zero modes which can only be soaked up 
with higher than four-derivative interactions.

For four gravitons the full instanton  corrections have been 
conjectured to take the form \cite{GG}
\be
{\cal{L}}_{R^4}=\frac{1}{3\cdot 2^7}
f_0(\tau,\btau)\left(t_8t_8+\frac{1}{8}\varepsilon_{10}
\varepsilon_{10}\right)R^4 \, , 
\ee
where $f_0(\tau,\btau)$ is the non-holomorphic modular form \cite{M} 
\be
f_0(\tau,\btau)={\sum_{m,n}}'\frac{\tau_2^{3/2}}{|m+n\tau|^3}\, , 
\ee
and the sum extends over integers $(m,n)\neq (0,0)$.  In addition, 
$f_0(\tau,\btau)$ has  the small $\tau_2^{-1}$ expansion 
\begin{eqnarray}
f_0(\tau,\btau)&=& 2\zeta(3)\tau_2^{3/2}+\frac{2\pi}{3}\tau_2^{-1/2}
+8\pi\tau_2^{1/2}\sum_{m\neq 0,n\geq 1}\left|\frac{m}{n}\right| e^{2i\pi 
mn\tau_1}K_1(2\pi|mn|\tau_2) \nonumber \\
&=&  
 2\zeta(3)\tau_2^{3/2}+\frac{2\pi}{3}\tau_2^{-1/2}\label{f0} \\
&& +4\pi^{3/2}\sum_{m,n\geq 1}\left(\frac{m}{n^3}\right)^{1/2}(e^{2i\pi mn\tau}
+e^{-2i\pi mn\btau})\left(1+\sum_{k=1}^{\infty}(4\pi mn\tau_2)^{-k}
\frac{\Gamma(k-1/2)}{\Gamma(-k-1/2)k!}\right)\, , \nonumber 
\end{eqnarray}
where $K_1$ is a Bessel function.  
This proposal for the exact $R^4$ corrections satisfies several consistency 
requirements:
\begin{enumerate}
\item[i)] 
$SL(2,\bf{Z})$ invariance. 
The function 
$f_0(\tau,\btau)$ is modular-invariant, so that ${\cal{L}}_{R^4}$ is invariant 
as well. 
\item[ii)]
It reproduces the correct perturbative expansion.
\item[iii)]
The non-perturbative corrections are of the expected form: there are only 
multiply-charged single D-instanton contributions.
\end{enumerate}

It should be noted that although the proposed form of the $R^4$ corrections to 
the effective action satisfy the above constraints, there is no proof 
for its validity. However, from the type IIA side \cite{GV,GGV},
as well as from lower dimensional compactifications of type IIB \cite{KP},
there exist strong arguments supporting this form of the $R^4$ terms. 
Nevertheless, it does not provide the full four-point Lagrangian since the 
analogous corrections to the other modes (antisymmetric fields and scalars)
 are lacking. One should expect 
that the substitution of the Riemann tensor $R_{MNPQ}$ with the modified 
one in eq.(\ref{hR}), as suggested by the tree-level result, is the full 
answer. However, in this case, the complex scalar and the 
antisymmetric fields are included in an non-modular-invariant 
way, which  explicitly breaks the 
conjectured $SL(2,\bf{Z})$ symmetry of the type IIB theory. 
We will construct  below a full $SL(2,\bf{Z})$-invariant effective action.

\section{ The $SL(2,\bf{Z})$-invariant type IIB effective action}

The proposed  $SL(2,\bf{Z})$-invariant four-point 
effective action compatible with the tree-level NS-NS sector,  which
includes the complex scalar and the antisymmetric two-form fields, is 
\begin{eqnarray}
{\cal{S}}&=&\frac{1}{2}\int d^{10}x\sqrt{-g} 
\left\{\mbox{\phantom{\huge{X}}}\right.\!\!\!\!\!\!\!\!\!\!
R-\frac{1}{2\tau_2^2}\partial_M\tau\partial^M\btau
-\frac{1}
{6}G_{KMN}\bar{G}^{KMN}
+\frac{1}{3\cdot 2^7}(t_8t_8+\frac{1}{8}\varepsilon_{10}\varepsilon_{10})\times
\nonumber \\ 
&&\left[\mbox{\phantom{\huge{X}}}\right.\!\!\!\!\!\!\!\!\!\!\frac{1}{2}
f_0(\tau,\btau)
\left(\mbox{\phantom{\huge{X}}}\right.\!\!\!\!\!\!\!\!\!\!R^4+
12R^2DP\,D\bar{P}-6RDP\,D\bar{G}^2 
+3R^2 DG\,D\bar{G}\nonumber \\
&&
+6DP^2\,D\bar{P}^2+\frac{3}{8}DG^2\,D\bar{G}^2
+6DP\,D\bar{P}\,DG\,D\bar{G}\!\!\!\!\!\!\!\!
\left.\mbox{\phantom{\huge{X}}}\right)
\nonumber \\
&&+f_1(\tau,\btau)\left(\mbox{\phantom{\huge{X}}}\right.\!\!\!\!\!\!\!\!\!\!
-4R^3DP +\frac{3}{2}R^2DG^2-12RDP^2\,D\bar{P}-6RDP\,DG\,D\bar{G} \label{f} \\
&&
+3DP\,D\bar{P}\,DG^2+\frac{3}{2}DP^2\,D\bar{G}^2+\frac{1}{4}
DG^3\,D\bar{G}\!\!\!\!\!\!\!\!
\left.\mbox{\phantom{\huge{X}}}\right)\nonumber \\
&&+
f_2(\tau,\btau)\left(\mbox{\phantom{\huge{X}}}\right.\!\!\!\!\!\!\!\!\!\!
6R^2DP^2-3RDPDG^2+4DP^3\,D\bar{P}+3DP^2\,DG\,D\bar{G}+\frac{1}{16}DG^4
\left.\mbox{\phantom{\huge{X}}}\!\!\!\!\!\!\!\!\!\!\right) \nonumber \\
&&
+f_3(\tau,\btau)\left(\mbox{\phantom{\huge{X}}}\right.\!\!\!\!\!\!\!\!\!\!
-4RDP^3+\frac{3}{2}DP^2DG^2 \!\!\!\!\!\!\!\!\!\!
\left.\mbox{\phantom{\huge{X}}}\right)
+f_4(\tau,\btau)
DP^4
+c.c. \!\!\!\!\!\!\!\!\!\!
\left.\mbox{\phantom{\huge{X}}}\right] \!\!\!\!\!\!\!\!\!\!
\left.\mbox{\phantom{\huge{X}}}\right\}\, , \nonumber  
\end{eqnarray}
where $DG$ stands for $(DG)_{MNPQ}\equiv D_{M}G_{NPQ}$, $DP$ for 
$(DP)_{MNPQ}\equiv g_{MP}D_NP_Q$ and similarly for $D\bar{G}$ and $D\bar{P}$
of $U(1)$ charge $q=-1,-2$, respectively. 
The functions $f_k(\tau,\btau)$ are defined as 
\be
f_k(\tau,\btau)=
{\sum_{m,n}}'\frac{\tau_2^{3/2}}{(m+n\tau)^{3/2+k}(m+n\btau)^{3/2-k}}\, , 
\label{fk}
\ee
for $k\in\bf{Z}$. They transform under  $SL(2,\bf{Z})$ as 
\be
f_k(\tau,\btau)\rightarrow \left(\frac{c\tau+d}{c\btau+d}\right)^k
f_k(\tau,\btau)\, , ~~~ \left(\matrix{a & b\cr c& d}\right)\in SL(2,\bf{Z})\, .
\ee 
These functions have the 
correct modular transformation properties to render the action
$SL(2,\bf{Z})$-invariant. Moreover, the effective action  we propose 
satisfies all the criteria listed in the previous section. Namely, it has only
tree-level and one-loop perturbative corrections in the NS-NS sector, which 
is an 
extension of the non-renormalization theorem of the $R^4$ term. 
This can be seen by examining the small $\tau_2^{-1}$ expansion of $f_k$,
which follows from 
\begin{eqnarray}
\left(k+2i\tau_2\frac{\p}{\p\tau}\right)f_k&=&\left(\frac{3}{2}+k\right)
f_{k+1} \, . \label{k1} 
\end{eqnarray}
As a result we obtain 
\be
f_k(\tau,\btau)=2\zeta(3)\tau_2^{3/2}+c_k\tau_2^{-1/2}+\cdots\, , \label{ck}
\ee
where 
\be
c_0=\frac{2\pi^2}{3}\, , ~~~\mbox{and} ~~~ c_{k+1}=\frac{2k-1}{2k+3}c_k\, , ~~~
k=0,1,...\, ,
\ee
and the dots in eq.(\ref{ck}) stand for instanton corrections. 

The action (\ref{f}) we propose has been constructed from 
the tree-level one (\ref{LL}) by replacing 
\be
\nabla_M\p_N\phi\rightarrow D_MP_N +D_M\bar{P}_N 
\ee
for the dilaton and 
\be
e^{-\phi/2}\nabla_MH^1_{NPQ}\rightarrow -\frac{i}{\sqrt{2}}D_MG_{NPQ}+
\frac{i}{\sqrt{2}}D_M\bar{G}_{NPQ} \label{HHH}
\ee
for the NS-NS antisymmetric tensor. Notice that when $\chi=0\, , H^2=0$,
the r.h.s. of eq.(\ref{HHH}) gives $e^{-\phi/2}\nabla_MH^1_{NPQ}+
e^{-\phi/2}\p_M\phi H^1_{NPQ}$, where the second term contributes to at least 
six-point amplitudes, which is not relevant to our discussion.  
Under modular transformations, each term in the resulting expression 
is multiplied by 
a factor $\left(\frac{c\tau+d}{c\btau+d}\right)^{q/2}$,  
which we finally compensate by replacing 
$2\zeta(3)\tau_2^{3/2}$ by $f_{q/2}(\tau,\btau)$. Notice that the terms 
linear or cubic in $\nabla H^1$, which vanish anyway at tree level 
in the expansion (\ref{LL}) because of the Bianchi identity would have 
implied  additional contributions in (\ref{f}) involving $f_{k/2}$'s with $k$ 
odd. However, it is easily seen from their definition in eq.(\ref{fk}) that
these functions vanish identically.

As a non-trivial check of our proposal, 
there exists a non-perturbative calculation of type IIB on $K3$ for the 
four-point amplitude involving two gravitons and two dilatons \cite{APT}.  
This term  corresponds to  
\be
A_{hh\phi\phi}\propto R^{ijkl}R_{ijkl}(\p_\mu\p_\nu\phi)(\p^\mu\p^\nu\phi)\, 
, \label{RRR} 
\ee
where the Latin and Greek indices refer to the internal $K3$ and to the
six-dimensional  spacetime, respectively.
 By taking the large $K3$ volume limit, 
one finds  that this amplitude is multiplied by 
the function $f_0(\tau,\btau)$ in 
ten dimensions. However, in our case from the action (\ref{f}) this 
amplitude is proportional to 
\be
\left(t_8^{ijkl\mu\nu\rho\sigma}t_8^{mnpq\kappa\lambda\varphi\omega}+
\frac{1}{8}\varepsilon_{10}^{ijkl\mu\nu\rho\sigma\alpha\beta}
{\varepsilon_{10}^{mnpq\kappa\lambda\varphi\omega}}_{\alpha\beta}\right)
R_{ijmn}R_{klpq}
g_{\nu\lambda}g_{\sigma\omega}\nabla_{\mu}\p_{\kappa}\phi
\nabla_{\rho}\p_{\varphi}\phi\, ,
\ee
and can be seen to vanish  by a straightforward calculation. 
This apparent paradox can be avoided\footnote{We would like to thank 
I. Antoniadis  and B. Pioline for their contribution on this point.}
by recalling that there exists an 
additional 
contribution to the six-dimensional  amplitude. This arises from
the the second and third terms in eq.(\ref{hR1})  
in the definition of $\hat{R}$ in eq.(\ref{hR}). In that case one obtains  
the additional contribution    
\begin{eqnarray}
&&\left(t_8^{ijkl\mu\nu\rho\sigma}t_8^{mnpq\kappa\lambda\varphi\omega}+
\frac{1}{8}\varepsilon_{10}^{ijkl\mu\nu\rho\sigma\alpha\beta}
{\varepsilon_{10}^{mnpq\kappa\lambda\varphi\omega}}_{\alpha\beta}\right)
R_{ijmn}R_{klpq}R_{\mu\nu\kappa\lambda}\times\nonumber \\&&
\mbox{\phantom{XXXXXXXXXXXXXXXXXXX}}
\left(\frac{1}{2}g_{\sigma\omega}\p_\rho\phi\p_\varphi\phi
- g_{\rho\varphi}g_{\sigma\omega}\p_\gamma\phi
\p^\gamma\phi\right) \, . \label{tt}
\end{eqnarray}
By  expanding around the background $g_{MN}=(\eta_{\mu\nu},
g_{mn}^{K3})$ and by partially integrating the two derivatives  
in  $R_{\mu\nu\kappa\lambda}$, we get a term proportional to the r.h.s. of 
eq.(\ref{RRR}). Finally, this term should be promoted in the exact 
$SL(2,\bf{Z})$-invariant Lagrangian to 
\be
 f_0(\tau,\btau)R^{ijkl}R_{ijkl}D_\mu P_\nu D^\mu\bar{P}^\nu \, , 
\ee
which  indeed reproduces the result of \cite{APT}.

\section{Conclusions}

We have conjectured here an S-duality invariant  effective four-point action 
of type IIB theory. The guiding principles we used were
basically the $SL(2,\bf{Z})$ invariance and the expectation 
that the perturbative 
corrections stop at one loop for the NS-NS sector. We found that these 
principles can be satisfied 
by using the functions $f_k(\tau,\btau)$ we introduced. They 
give the correct perturbative corrections and the non-perturbative ones 
are of the expected form. Moreover, the vanishing of four-point amplitudes 
that involve one or three antisymmetric fields is consistent, as we have 
discussed, with the vanishing of the $f_k$ forms for half-integer 
$k$. Finally,
the result of \cite{APT} is consistent with our proposal since, for the 
$K3$ compactification they consider, it arises 
from a five-point term in ten dimensions. 
Finally, a simple test of our proposal would be an explicit 
determination of the one-loop coefficients $c_k$ defined in eq.(\ref{ck}).   

\vspace{.5cm}

\noindent
{\bf{Acknowledgement}} 
\vspace{.3cm}

We would like to thank I. Antoniadis, E. Kiritsis, 
C. Kounnas, N. Obers and B. Pioline  for very  useful discussions.     
H.P. would like to thank the kind hospitality of T.U. Munich  where 
this work was initiated.

\end{document}